\documentclass[graybox]{svmult}
\usepackage{mathptmx}       
\usepackage{helvet}         
\usepackage{courier}        
\usepackage{makeidx}         
\usepackage{graphicx}        
\usepackage{multicol}        
\usepackage[bottom]{footmisc}

\usepackage[utf8x]{inputenc}
\makeindex             

\begin{document}

\title*{Localized structures in broad area VCSELs: experiments and delay-induced motion}
\author{Mustapha Tlidi, Etienne Averlant, Andrei Vladimirov, Alexander Pimenov,
Svetlana Gurevich and Krassimir Panayotov}
\authorrunning{M. Tlidi, E. Averlant, A. Vladimirov, A. Pimenov, S. Gurevich and K. Panayotov}
\institute{Mustapha Tlidi \at Faculté des Sciences de l'Université libre de Bruxelles, Belgium 
\email{mtlidi@ulb.ac.be}
\and Etienne Averlant \at Faculté des Sciences de l'Université libre de Bruxelles, Belgium  and IR-TONA, 
Vrije Universiteit Brussel, Belgium \email{eaverlan@ulb.ac.be}\and Andrei Vladimirov \at Weierstrass Institut for applied physics und 
stochatics, Berlin, Germany \email{vladimir@wias-berlin.de}\and Alexander Pimenov \at Weierstrass Institut 
for applied physics und stochatics,Berlin, Germany \email{Alexander.Pimenov@wias-berlin.de}\and Svetlana 
Gurevich \at Institut für theoretische Physik, Münster, Germany \email{gurevics@uni-muenster.de}\and Krassimir Panayotov \at IR-TONA, Vrije 
Universiteit Brussel, Belgium and Institute of solid State physics, Sofia, Bulgaria 
\email{kpanajot@b-phot.org}}
\maketitle
\abstract{We investigate the space-time dynamics of a Vertical-Cavity Surface-Emitting Laser (VCSEL) subject to optical injection
 and to delay feedback control. Apart from their technological advantages, broad area VCSELs allow creating localized light 
 structures (LSs). Such LSs, often called Cavity Solitons, have been proposed to be used in information 
 processing, device characterization, and others. After a brief description of the experimental setup, we present experimental 
 evidence of stationary LSs. We then theoretically describe this system using a mean field model. We perform a real order 
 parameter description close to the nascent bistability and close to large wavelength pattern forming regime. We theoretically 
 characterize the LS snaking bifurcation diagram in this framework.  The main body of this chapter is devoted to theoretical 
 investigations on the time-delayed feedback control of LSs in VCSELs. The feedback induces a spontaneous motion of the LSs, 
 which we characterize by computing the velocity and the threshold associated with such motion. In the nascent bistability 
 regime, the motion threshold and the velocity of moving LSs depend only on the feedback parameters. However, when considering 
 the previously introduced mean-field model, theoretical predictions indicate that both motion threshold and velocity are 
 strongly affected by the phase of the delay and by the carrier relaxation rate.
}
\section{Introduction}
\label{intro}
Spontaneous symmetry breaking and self-organization phenomena have been observed in various fields of nonlinear science such as 
nonlinear optics, fibre optics, fluid mechanics, granular matter or plant ecology. The link between the well known Turing 
instability and transverse patterns formation in nonlinear optics was established for the first time by Lugiato and Lefever 
\cite{LuLe}. In their seminal paper, they considered an optical resonator filled with a passive nonlinear medium and driven by a 
coherent radiation beam. Since then, many driven systems have proven to allow periodic patterns in the transverse section of 
their output beam. Besides a periodic distribution of light, LSs may form in the plane perpendicular to the 
propagation axis. They are often called localized spots and localized patterns, or cavity solitons which appear either isolated, 
randomly distributed or self-organized in clusters forming a well-defined spatial pattern \cite{k32,Scroggie}.  When LSs are 
sufficiently separated from each other, localized peaks are independent and randomly distributed in space. However, when the 
distance between peaks decreases they start to interact via their oscillating, exponentially decaying tails 
\cite{VladimirovPRE02,TlidiIQE,Tli-Lef-Vla,Turaev12}. LSs have been reported in  nonlinear resonators such as lasers with 
saturable absorbers \cite{k29,Rev2,k31}, in passive nonlinear resonators \cite{k32,Scroggie,k59-1}, optical parametric 
oscillators \cite{k33,k34}, in left-handed materials \cite{Kockaert06,Gelens07,kozyrev,TlidiPRA11}, in exciton-polariton 
patterns in semiconductor microcavities \cite{EgorovPRB14-1,EgorovPRB14-2}  and in the framework of the Ginzburg-Landau equation 
\cite{Brand89,MalomedPRA91,MalomedPRE98,BessePRE13,SKARKAPRA14}. Phase solitons have been demonstrated far from any pattern 
forming instability \cite{k40,k40-1,k41,k42,k52-1,OtoPRL13}.

Localized structures are homoclinic solutions (solitary or stationary pulses) of partial differential equations. The conditions 
under which LSs and periodic patterns appear are closely related. Typically, when the Turing instability becomes subcritical,
there exists a pinning domain where LSs are stable. This is a universal phenomenon and a well-documented issue in various fields
of nonlinear science. The experimental observation of LSs in driven nonlinear optical cavities has further motivated the interest
in this field of research. In particular, LSs could be used as bits for information storage and processing. Several overviews
have been published on this active area of research
\cite{Rev1,Rev2,Mandel1997,Rev3,Rev4,Rev5,Rev6,Rev7,Rev8-1,Rev9,Rev10,Rev12,Rev13,Rev14,Rev15,Rev16,Rev18,Rev19,Rev20}.

Many theoretical and experimental studies on LS formation in VCSELs have been realized \cite{Taranenko_pra00,Pedaci_apl08b}.
They have been experimentally observed in broad area VCSELs both below \cite{Taranenko_pra00,Barland_n02} and above 
\cite{Hachair_jstqe06} the lasing threshold when injecting a holding beam with appropriate frequency and power. A spatially LS 
has also been found in a medium size VCSEL, but only by using its particular polarization properties \cite{Hachair_pra09}. 
Cavity soliton lasers (CSLs) in a VCSEL system without a holding beam have been demonstrated both experimentally
\cite{Tanguy_prl08} and theoretically \cite{Paulau_pre08} in VCSELs subject to frequency selective optical feedback and in face
to face coupled VCSELs \cite{Genevet_prl08,Columbo_ejpd10}. In these systems, the VCSELs are placed in self imaging optical
systems with either an external grating or another VCSEL biased below lasing threshold, so that the system becomes bistable.
Lasing spots spontaneously appear in these systems and can be switched on and off by another laser beam. As a matter of fact
a broad area VCSEL with saturable absorber has been the first system in which LSs have been predicted and studied theoretically
\cite{Rosanov_os88,Vladimirov_job99,Fedorov_pre00}. LSs in a monolithic optically pumped VCSEL with a saturable absorber have
been demonstrated in \cite{Elsass} and their switching dynamics studied in \cite{Elsass_apb10}. Several applications of LSs in
VCSELs have been demonstrated: optical memory \cite{Pedaci_apl06}, optical delay line \cite{Pedaci_apl08} and optical microscopy
\cite{Pedaci_apl08b}.

In this chapter, we investigate the formation of LSs in Vertical-Cavity Surface-Emitting Lasers (VCSELs) subject to both optical 
injection and delay feedback control. These lasers are characterized by a large Fresnel number and a short cavity. VCSELs are 
the best candidate for localized structures formation in their transverse section. The first VCSELs were fabricated in 1979 
\cite{Soda_jjap79} and later on they reached performances comparable to those of edge-emitting lasers \cite{Iga03,VCSELs13}. 
Nowadays VCSELs are replacing edge-emitting lasers in short and medium distance optical communication links thanks to their 
inherent advantages: much smaller dimensions, circular beam shape that facilitates coupling to optical fibres, two-dimensional 
array integration and on wafer testing that brings down the production cost \cite{VCSELs13}. As VCSELs emit light perpendicular 
to the surface and the active quantum wells, their cavity length is of the order of 1 $\mu$m - the wavelength of the generated 
light. Thanks to the maturity of the semiconductor technology VCSELs can be made homogeneous over a size of hundreds of $\mu$ms 
while the characteristic LS size is about 10 $\mu$m.  The timescales of the semiconductor laser dynamics and LS formation are in 
the ns scale, which allows for fast and accurate gathering of data. Finally, VCSEL physics and dynamics are quite well 
understood \cite{VCSELs13,Panajotov_jstqe13,Panajotov_jqe09}.

We investigate experimentally and theoretically the formation of stationary LSs in VCSELs and we describe theoretically the 
effect of a time delayed feedback on the stability of LSs. For this purpose, we adopt the Rosanov-Lang-Kobayashi approach for 
modelling of delay feedback \cite{Rosanovdelay,RLang1980}.   

This chapter is organized as follows. After an introduction, we provide a description of the VCSELs in Sec (2), the experimental
setup and the observation of localized structures in a medium size VCSEL are described in Sec. (3). A theoretical description 
based on mean field model and the derivation of the generalized Swift-Hohenberg equation are given in Sec. (4). Stationary LSs 
and their bifurcation diagram are presented in Sec. (5). LSs brought into motion under the effect of delay feedback are 
discussed  in Sec. (6). We conclude and draw some perspectives of the present work in Sec. (7).

\section{Vertical-Cavity Surface-Emitting Lasers}
\label{vcsel}
The structure of a VCSEL is by far more complex than the one of an edge-emitting semiconductor laser. But
this complexity, and the inherent fabrication costs, did not keep VCSELs from becoming the second most
produced laser type \cite{Iga-Review,Iga-Review2,VCSEL-Michalzik}. The biggest advantages of this structure
are the circular emission pattern and low divergence, which allow to easily couple the light to an optical
fibre for optical interconnects.  Moreover, the high reflectivity of the Bragg mirrors provide a low lasing
threshold, which makes  VCSELs the choice lasers for low power applications (such as optical mice for
example). For high power applications, VCSELs can easily be coupled into arrays \cite{Lasers}, which
provides a high power coherent low divergence beam.

\begin{figure}[t]
\sidecaption[t]
\centering
\includegraphics[width=0.6\textwidth]{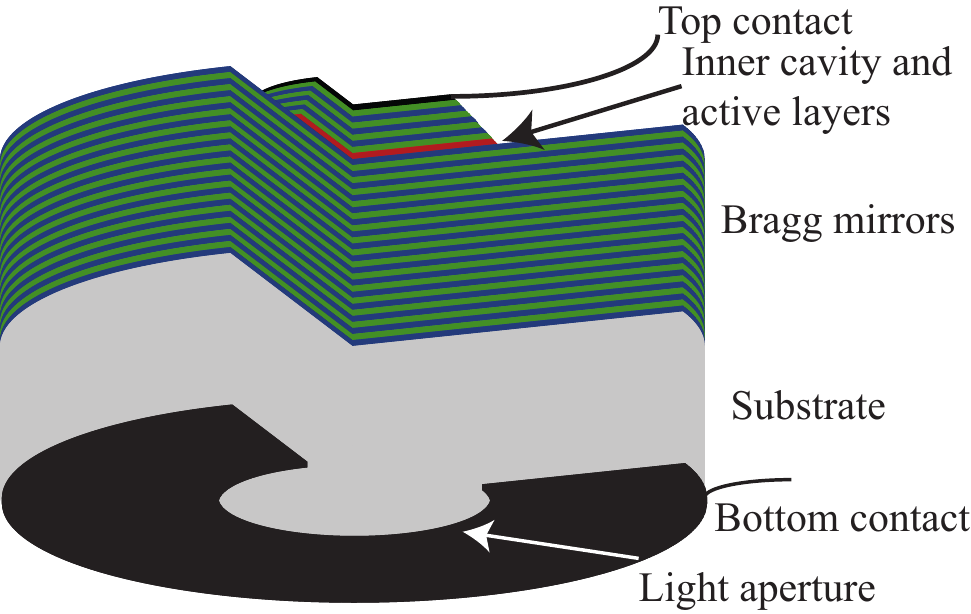}
\caption{View of a bottom-emitting VCSEL. The blue and green layers constitute the pairs in the Bragg mirrors. The red
element is the cavity containing the quantum wells. The grey part is the substrate on which the structure has been grown and 
through which the light is emitted.}
\label{bottom-vcsel}
\end{figure}

The emission surface of a VCSEL can be made from a few $\mu$m to some 200 $\mu$m - the first ones are
best suited for low power applications, whereas the second category is typically used for spectroscopy.
Apart of their use in spectroscopy, broad area VCSELs are best suited for LSs studies.
They have several characteristics that makes them a choice material:
\begin{itemize}
\item Their use in information technologies provides a mastering of the fabrication process, leading to à la carte properties;
\item The response time is much faster than the one in liquid crystals or photorefractive media, which makes them more likely to 
be used in potential high speed applications;
\item Experiments can be carried out on a single one hundred micrometer chip, in comparison with some meter needed for creating 
LSs in gas cells;
\end{itemize}
The broad-area bottom emitting VCSEL structure we use in our experiments is described in Fig. \ref{bottom-vcsel}. The top and 
bottom distributed Bragg reflectors consist of 30 and
20.5 Al$_{0.88}$Ga$_{0.12}$As–GaAs layer pairs, respectively. The active
region consists of three In$_{0.2}$Ga$_{0.8}$As quantum wells embedded in GaAs barriers and AlGaAs cladding
layers\cite{Lasers}. 
The bottom-emitting configuration allows the stand-alone VCSEL to have a better (more homogeneous) current
distribution in the transverse plane and, therefore, is more suitable for producing LSs. Moreover
a heat sink can be directly mounted on top of the $p$-contact and the $p$-doped DBR, which is the main source of heat. The 
temperature distribution is hence more homogeneous as well.

\section{Experimental observations}

\subsection{Description of the experimental setup}
\begin{figure}[t]
\sidecaption[t]
\centering
\includegraphics[width=0.6\textwidth]{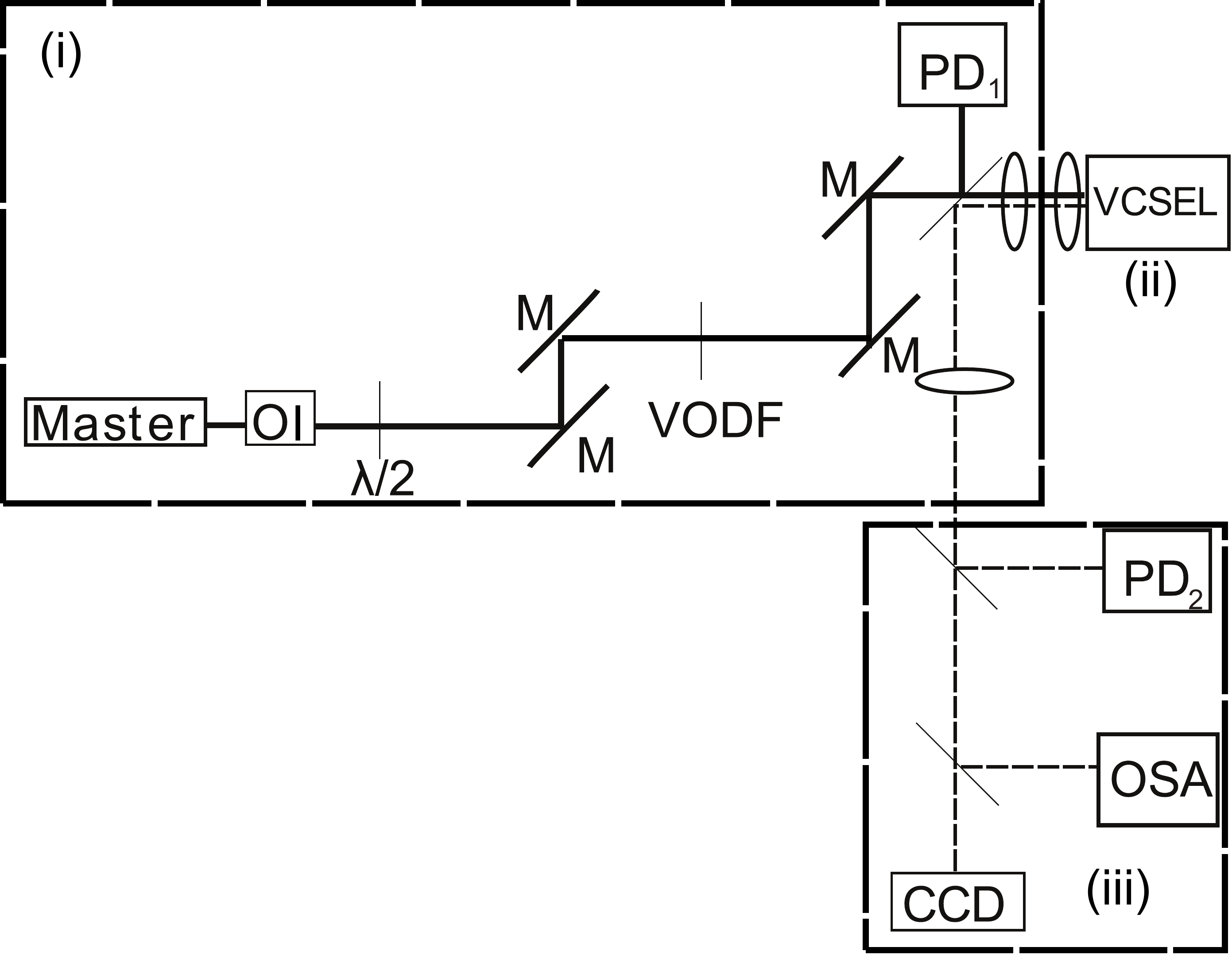}
 \caption{Experimental setup schematic. The full line is the path of the light from the master laser,
 whereas the dashed line is the path followed by the light from the VCSEL. (i): injection preparation and
 monitoring; Master: master laser, OI: optical isolator, $\lambda/2$ : half wave plate, M: mirror, VODF:
 variable optical density filter; (ii) : VCSEL; (iii): analysis branch; PD: photodiode, OSA: optical
 spectrum analyser. Reprinted from \cite{Averlant2014}}
\label{scheme}
\end{figure}

The experimental setup used for the generation of two-dimensional LSs is shown in Fig. \ref{scheme}.
The injection part consists of a tunable External Cavity Diode Laser (Master) in a Littrow configuration, optically isolated 
from the rest of the experiment. The linear polarization of the injected light is tuned to match the one of the VCSEL. In order 
to produce LSs, both the optical injection power and the detuning between the injection frequency and the VCSEL resonance 
frequency are adjusted.

The VCSEL we use is a $80\mu$m diameter, bottom emitting InGaAs multiple quantum well VCSEL, as described in Sec.(\ref{vcsel}) 
which has a threshold current of $42.5$ mA at $25.0^{\circ}$C.

We analyse the optical spectrum, near field and output power of the VCSEL.

\begin{figure}[t]
\centering
\includegraphics[width=0.8\textwidth]{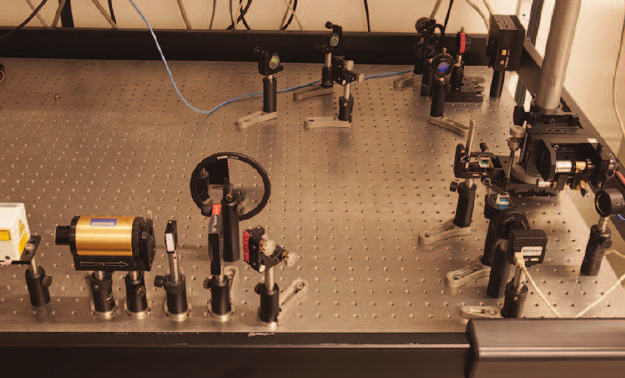}
\caption{Photography of the experimental setup. Left down side of the picture: injection preparation with
external cavity laser diode, optical isolator, half-wave plate and variable optical density filter.
Top right: injection monitoring and VCSEL. Down right: analysis branch; optical spectrum analyser, CCD
camera and photodiode.}
\label{Photo}
\end{figure}

A photography of the actual setup is shown in Fig.\ref{Photo}.
\subsection{Experimental observations}
\label{CS}

\begin{figure}[htbp]
\centering
\includegraphics[width=0.8\textwidth]{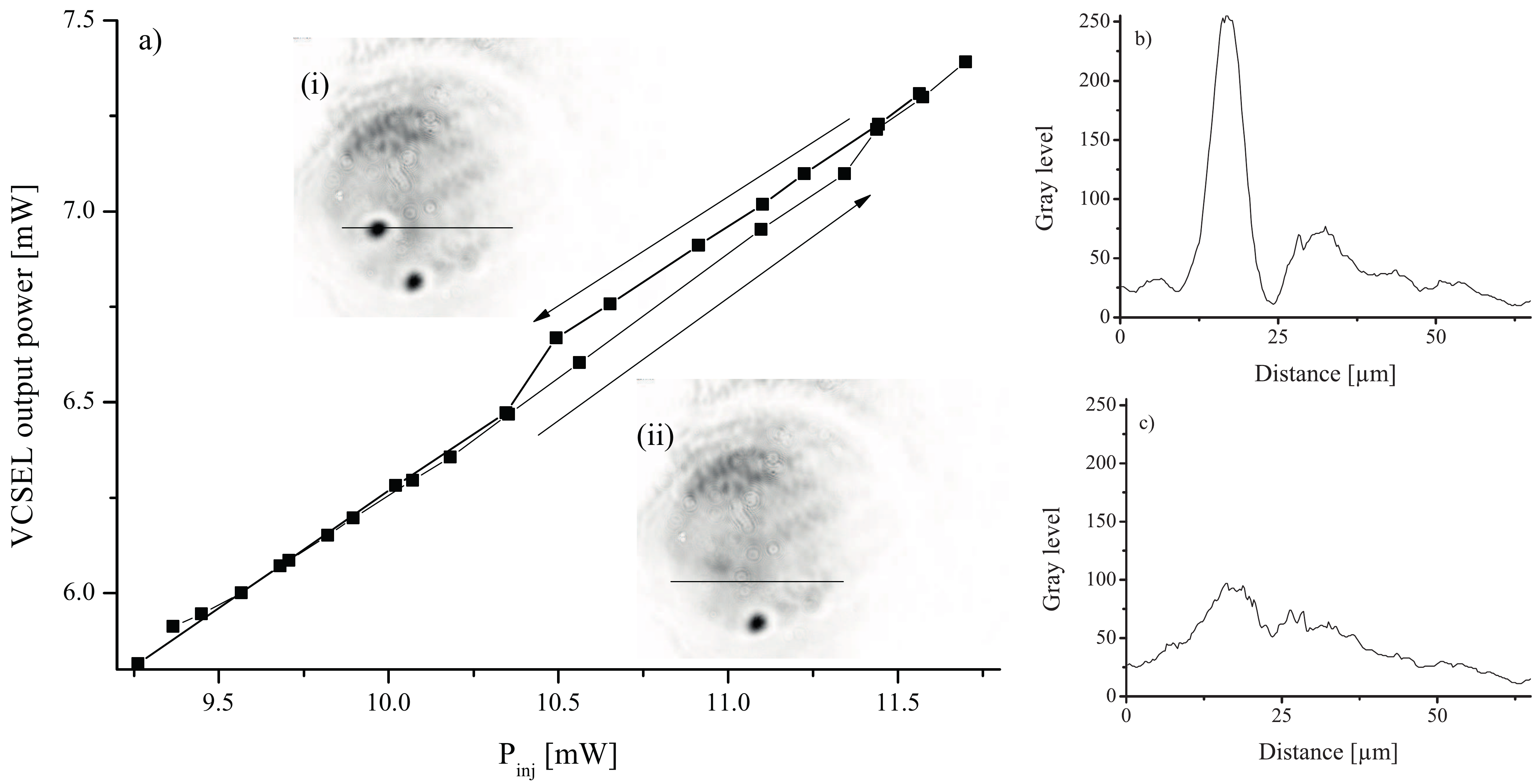}
 \caption{Bistability between one and two-peaked LSs inside the near field of the VCSEL as a function of
 the optical injection power. (a): power emitted by the VCSEL as a function of the optical injection power
 for $\theta=-174$GHz and a beam waist of $100\mu$m. The insets (i) and (ii) respectively represent near
 field profiles on the higher and lower branch of the hysteresis curve. (b) and (c): one dimensional profiles
 along the horizontal line drawn on the aforementioned insets. Redrawn from \cite{Averlant2014}}.
\label{bistab}
\end{figure}

\begin{figure}[htbp]
\centering
\sidecaption
\includegraphics[width=0.6\textwidth]{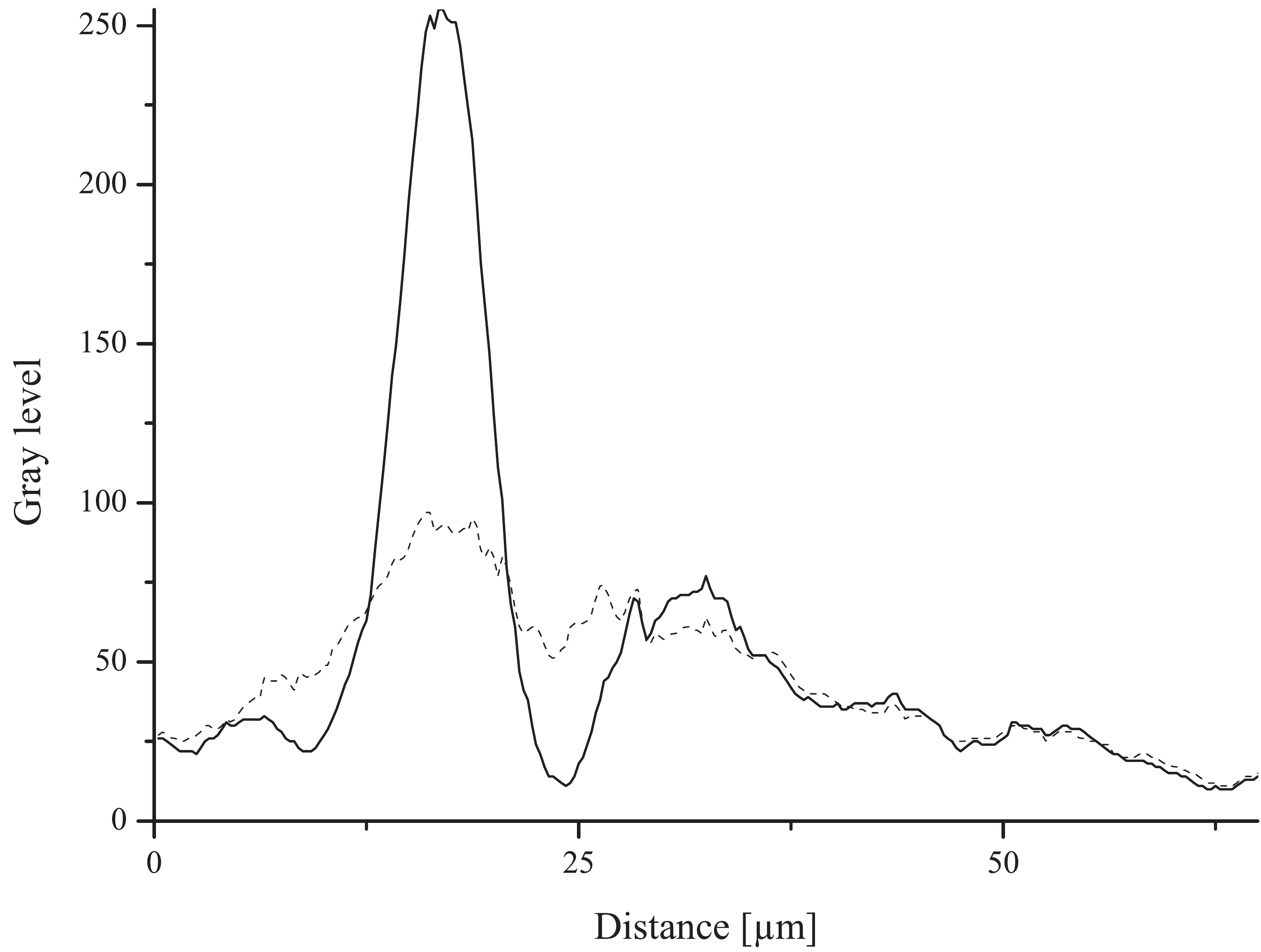}
 \caption{Cross sections along the solid lines indicated in Fig.\ref{bistab}(a), (i) and (ii). The dashed
 line is the state (ii)(lower branch of the hysteresis), whereas the full line is the system with a LS
 (upper branch of the hysteresis). Redrawn from \cite{Averlant2014}}
\label{lines}
\end{figure}

Tuning the width of the injection beam to 100 $\mu$m, the detuning to $-174$GHz, the VCSEL driving current
to $45.013$mA and its substrate temperature to $25.01^\circ$C, we obtain the bistability curve depicted in
Fig. \ref{bistab}. We continuously varied the optical injection power by increasing it, before decreasing it. The hysteresis 
phenomenon associated with this experiment is evidenced  in Fig.\ref{bistab}a). The insets represent near
 field profiles on the higher and lower branch of the hysteresis curve. One dimensional profiles along the horizontal lines 
 drawn on the insets are shown in Fig.\ref{bistab}b)(up) and c)(down). The lower branch corresponds to the "homogeneous" steady 
 state. In Fig.\ref{lines}, the difference between these two states is evidenced. On the higher branch, there is not only a 
 peak, but a clearly noticeable oscillating tail around this peak, which is characteristic of a LS.

\section{Mean field model}
In this section, we describe the dynamics of a VCSEL submitted to optical injection and delayed feedback. To do so, we assume an 
external cavity (i) in which the diffraction is compensated (ii) much longer than the 
characteristic diffraction length of the field. We further apply (i) a paraxial approximation (ii) a slowly varying envelope 
approximation. The mean field model describing the space-time 
evolution of the electric field envelope $E$ and  the  carrier density $N$ in a VCSEL subjected to optical injection and 
time-delayed feedback is then given by the following set of dimensionless partial differential equations
\begin{eqnarray}
\label{eq:dEdt}\frac{\partial E}{\partial t} &=& -\left(\mu+i\theta\right)E + 2C(1-i\alpha)(N-1)E   \\
&+& E_{i} -\eta e^{i\varphi} E(t-\tau)+ i\nabla^{2} E\,, \nonumber \\
\label{eq:dNdt} \frac{\partial N}{ \partial t} &=& -\gamma \left[ N -I + (N-1)\left|
E\right| ^{2} - d\nabla^{2} N\right] .
\end{eqnarray}
The parameter $\alpha$ describes the linewidth enhancement factor, $\mu$ and $\theta$ are the cavity decay rate and the cavity 
detuning parameter, respectively. The parameter $E_{i}$ is the amplitude of the injected field which we assume to be positive in 
order to fix the origin of the phase.  $C$ is the bistability parameter, $\gamma$ is the carrier decay rate, $I$ is the 
injection current, and $d$ is the carrier diffusion coefficient. The light diffraction and the carrier diffusion are described 
by the terms $i\nabla^{2}E$ and $d\nabla^{2}N$, respectively, where $\nabla^{2}$ is the Laplace operator acting in the 
transverse plane $(x,\,y)$. Below we consider the case when the laser is subjected to coherent delayed feedback from an external 
mirror. To minimize the effect of diffraction on the feedback field we assume that the external cavity is self-imaging 
\cite{Tlidi09}. The feedback is characterized by the delay time $\tau=2L_{ext}/c$, the feedback rate $\eta\ge 0$, and phase 
$\varphi$, where $L_{ext}$ is the external cavity length, and $c$ is the speed of light. The link between dimensionless and 
physical parameters is provided in \cite{Panajotov10}. Using the expression for the feedback rate 
$\eta=r^{1/2}(1-R)/R^{1/2}\tau_{in}$ given in \cite{Tartwijk}, where $r$ ($R$) is the power reflectivity of the feedback (VCSEL 
top) mirror and $\tau_{in}$ is the VCSEL cavity round trip time, we see that the necessary condition for the appearance of the 
soliton drift instability $\eta\tau>1$ \cite{Tlidi09} can be rewritten in the form 
$r>\frac{R \tau_{in}^2}{(1-R)^{2}\tau^2}$. In particular, for $R=0.3$ and $\tau=20\tau_{in}$ the latter inequality becomes 
$r>1.5\cdot 10^{-3}$.

To reduce the number of parameters, we introduce the following change of variables: $n=[2C(N-1)-1]/2$ and $e=E^*/\sqrt{2}$. The 
model Eqs. (\ref{eq:dEdt},\ref{eq:dNdt}) of a VCSEL driven by an injected field $Y=E_i/(2\sqrt{2})$ take the following form:
\begin{eqnarray}
\partial _{t}e &=& i\theta'e+(1+i\alpha)ne+Y +\eta'  e^{-i\psi}e(t-\tau)- i\nabla^{2} e, \label{eqreduced1}\\
\partial _{t}n &=& \gamma [P -n - (1+2n)|e| ^{2} + D\nabla^{2} n]. \label{eqreduced2}
\end{eqnarray}
The pump parameter $P$ is $P=C(I-1)-1/2$, $\gamma=\gamma'/2$, $D=2d$, $\eta'=\xi/2$, and $\theta'=(\theta+\alpha)/2$. The new time and space scales are $(t,\tau)=2(t',\tau')$ and $ \nabla^{2}_{\perp}=2 \nabla'^{2}_{\perp}$. Let us assume for simplicity that the detuning is $\theta'=0$ and the feedback phases are  $\psi =0$ or $\psi=\pi$.

The homogeneous steady states of Eqs. (\ref{eqreduced1},\ref{eqreduced2}) are  $Y=-e_s(1+i\alpha)(P-|e_s|^2)/(1+2|e_s|^2)$ and 
$n_s=(P-|e_s|^2)/(1+2|e_s|^2)$. We explore the vicinity of the nascent optical bistability regime where there exists a second 
order critical point marking the onset of a hysteresis loop.  The critical point associated with bistabilty is obtained when 
the output intensity as a function of the injection parameter $Y$ has an infinite slope, i.e.,  
$\partial Y/\partial |e_s|=\partial^{2} Y/\partial |e_s|^{2}=0$.  The coordinates of the critical point  are $e_c=(1-i\alpha)\sqrt{3/2(1+\alpha^2)}$, $n_c=-3/2$, $P_c=-9/2$, $D_c=8\alpha/[3(1+\alpha^2)]$ and 
$Y_c=(3/2)^{(3/2)}(1+\alpha^2)^{1/2}$.  Our objective is to determine a slow time and slow space amplitude equation which is valid under the following approximations: (i) close to the onset of bistabilty (ii) close to large wavelength symmetry breaking instability. Starting from Eqs. (\ref{eqreduced1}, \ref{eqreduced2}), the deviation $u$ of the electric field from its value at the onset of bistablity is shown to obey \cite{Tlidi12}
\begin{eqnarray}
 \label{eqSHE}
\partial _{t}u &=& y-u(p+u^2)+ \eta u(t-\tau) \label{SHeq}\\ \nonumber
&+&(d-\frac{5u}{2})\nabla^{2} u
-a \nabla^{4} u-2(\nabla u)^2,
\end{eqnarray}
where $a= (1-\alpha^2)/(4\alpha^2)$. The parameter $y$ denotes the deviation of the injected field amplitude from $Y_C$. The 
real variable $u$, the parameters $p$ and $d$ are the deviations of the electric field, the pump parameter and the carrier 
diffusion coefficient from their values at the onset of the critical point, respectively. In the absence of delay; i.e., 
$\eta=0$, Eq.(\ref{eqSHE}) is the generalized Swift-Hohenberg equation that has been derived for many far from equilibrium 
systems \cite{PRE-Kozyreff03,k8-3,Clerc2005}. Even in absence of the delay feedback term, the terms $u\nabla^{2}u$ and 
$(\nabla u)^2$ render Eq.(\ref{eqSHE}) nonvariational.

\section{Stationary localized structures}
LSs are nonlinear bright or dark peaks in spatially extended systems. Such structures have been observed 
in the transverse section of coherently driven optical cavities, and are often called cavity solitons. Currently they attract 
growing interest in optics due to potential applications for all-optical control of light, optical storage, and  information 
processing \cite{k32,Scroggie}. When they are sufficiently separated from each other, localized peaks are independent and 
randomly distributed in space. However, when the distance between peaks decreases they start to interact via their oscillating, 
exponentially decaying tails. This interaction then leads to the formation of clusters 
\cite{VladimirovPRE02,TlidiIQE,Tli-Lef-Vla,Turaev12}.  Mathematically speaking,  LSs are homoclinic solutions (solitary or 
stationary pulses) of partial differential equations. The conditions under which LSs and periodic patterns appear are closely 
related. Typically, when the Turing instability becomes sub-critical, there exists a pinning domain where localized structures 
are stable.  This is a universal  phenomenon  and a well documented issue in various fields of nonlinear science, such as 
chemistry, plant ecology,  or optics (see some recent overviews on this multidisciplinary issue (\cite{Rev12,Rev13,Rev20}). 

In this section we  describe some basic properties of stationary LS and their bifurcation diagrams in a one dimensional setting. 
In the absence of delay feedback, Eq. (\ref{SHeq}) admits a variety of LSs. The generalized Swift-Hohenberg equation(\ref{SHeq}) 
is numerically integrated using a classical spatial finite-difference method with forward temporal Euler integration. The 
boundary conditions are periodic in both transverse directions and the initial condition consists of a large amplitude peaks 
added to the unstable homogeneous steady state. We fix all the parameters except the amplitude of the injected field $y$. 
Examples of localized clusters having an odd or an even number of peaks are shown in Figs. 1b and 1c. They are 
obtained for the same parameter values and differ only by the initial condition.

\begin{figure}[htbp]
\centering
\includegraphics[width=1.0\textwidth]{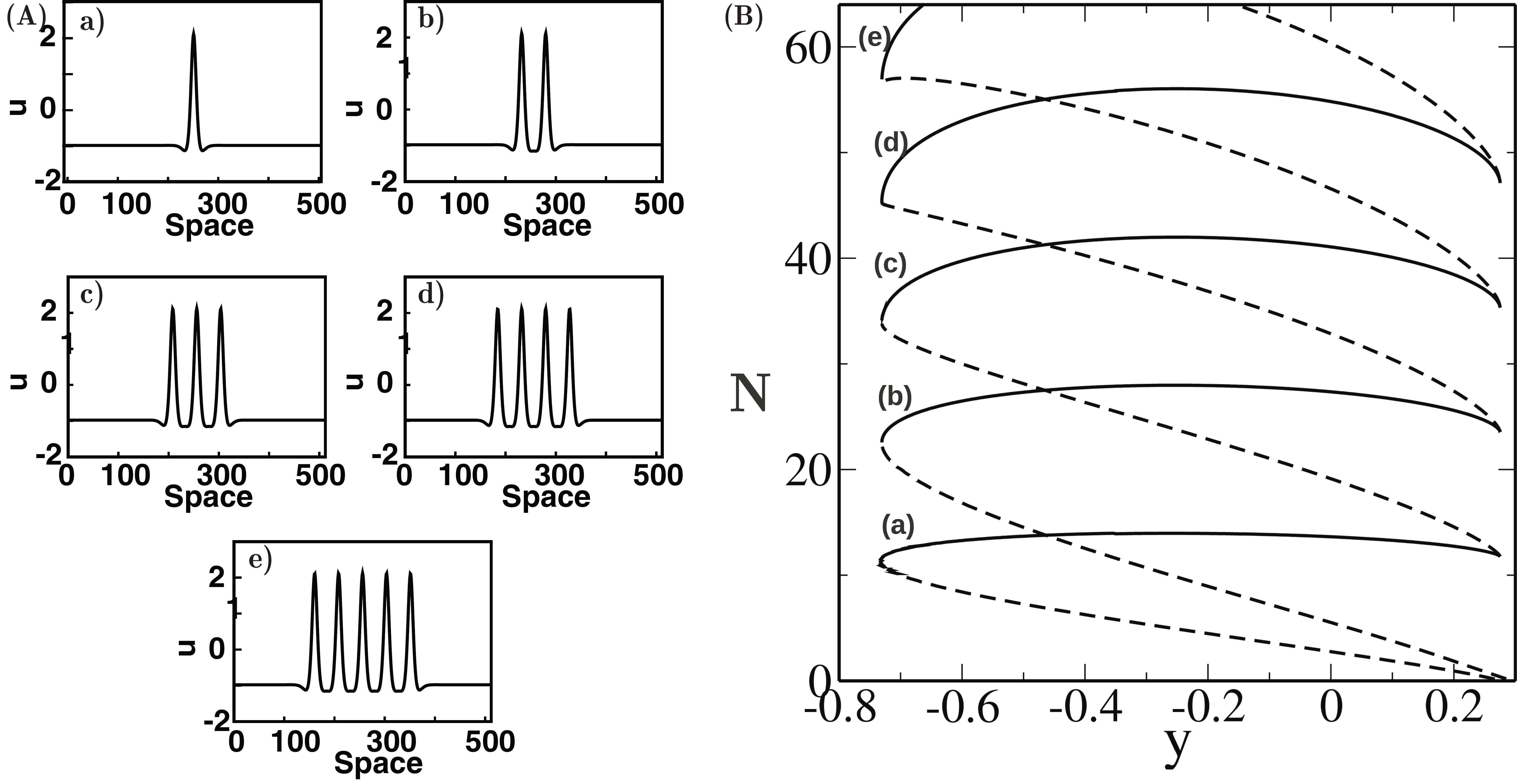}
 \caption{One dimensional LSs. (A) Multiple peaks localized structures with  odd or even number of peaks obtained for $y=-0.35$.
 (B) Snaking bifurcation diagram showing two inter-weaved snaking curves: the branches (a), (b), (c), (d) and (e) correspond to 
 states having 1, 2, 3, 4 and 5 peaks in fig. \ref{multiLS}A. Redrawn from \cite{Tlidi12}.
The full and the broken lines correspond, respectively, to stable and unstable localized branches of LSs. The parameters are $p=-0.7$, $d=-1.2$, $a=0.75$, and $\tau=\eta=0$.}
\label{multiLS}
\end{figure}

We examine the case of one-dimensional monostable system in the subcritical regime where the homogeneous steady state coexists 
with a spatially periodic structure. In addition, the system exhibits a high degree of multistability in a finite range of $y$ values often called the pinning region \cite{Pomeau}. More precisely, Eq. (\ref{SHeq}) then admits an infinite set of odd and even cavity solitons as shown in Fig. \ref{multiLS}A, i.e., a set of stationary solutions that exhibit $n = 2p-1$ or $n = 2p$ peaks, where $p$ is a positive integer. The limit $p \rightarrow \infty $, corresponds to the infinitely extended periodic pattern distribution.  In the pinning region, the width of LS is close to the half of the periodic wavelength structure. Since the amplitudes of localized patterns having different number of peaks are close to one another, in order to visualize the clusters properties, it is convenient to plot the ``$L_2$ -norm'' 
\begin{equation}
{\mathcal{N}}=\int dxdy|u-u_{s}|^{2}
\end{equation} 
as a function of the injected field $y$. This yields the two snaking curves with odd or even number of peaks as shown in 
bifurcation diagram of  Fig.  (\ref{multiLS}B). As ${\mathcal{N}}$  increases, at each turning point where the slope becomes infinite, a pair of additional peaks appears in the cluster. One sees that this behavior, referred to as homoclinic snaking phenomenon \cite{Champneys98,Coullet00,Clerc-Falcon,Knobloch06,Knobloch07,Tlidi-Gelens,VLT11}, and recently observed experimentally \cite{Haudin-Bortolozzo,VECSLEexp5}, corresponds to back and forth oscillations inside the pinning region. In two spatial dimensions,  the variety of stable localized patterns is much larger than in one dimensional system.

\section{Moving localized structures}
\begin{figure}[htbp]
\centering
\includegraphics[width=0.8\textwidth]{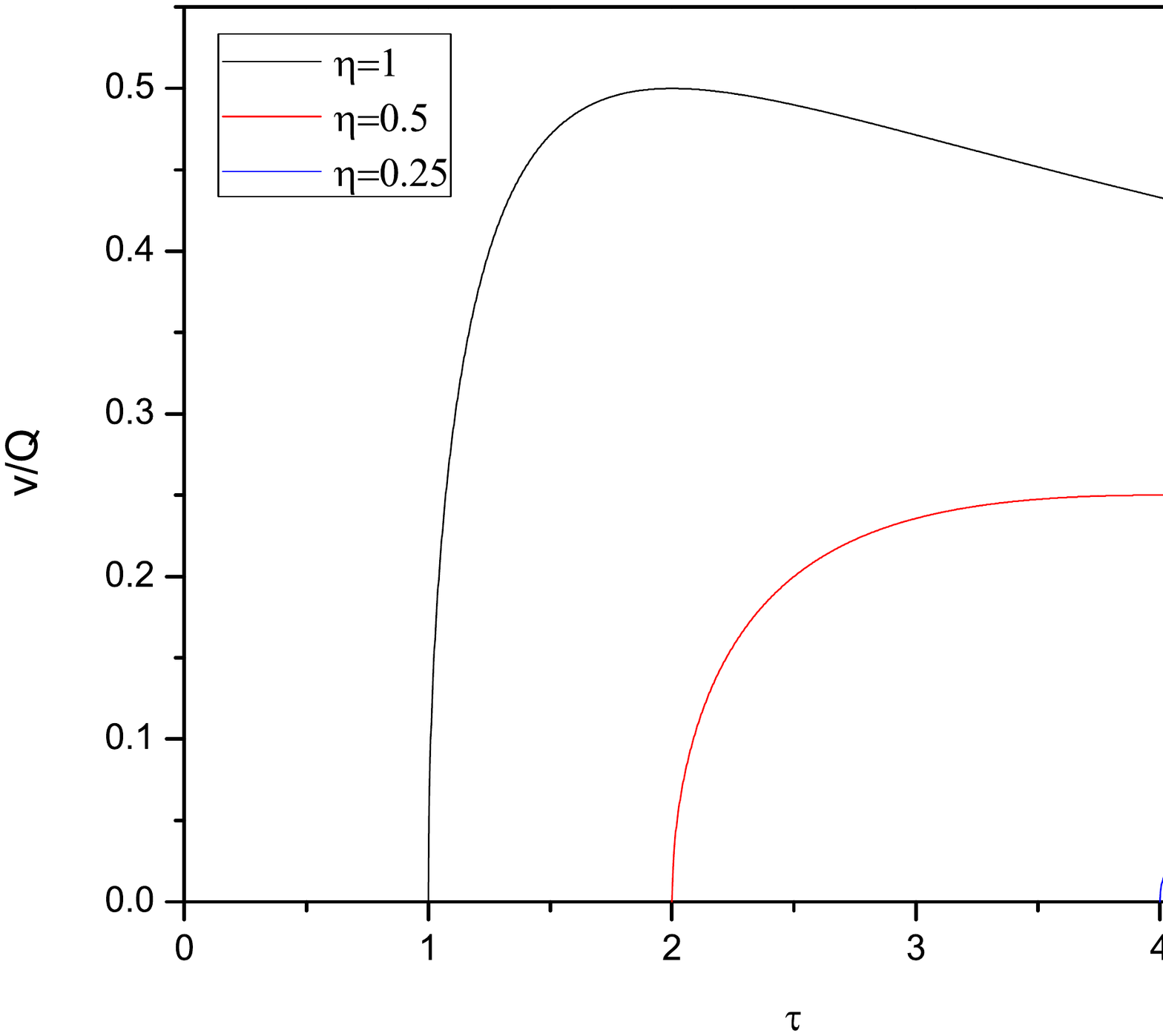}
 \caption{Velocity by unit of the factor $Q$ of moving localized structures as a function of the time delay $\tau$ for different 
 values of the delayed feedback strength $\eta$.}
\label{velocityGSH}
\end{figure}

In this section we investigate the effect of a time-delayed feedback control on the stability of LSs in VCSELs. This delayed 
feedback loop compares the electric field at the current moment of time and its values at some time in the past. Recent studies
that combined analytical and numerical analysis of the two-dimensional Swift–Hohenberg equation suggested that steady LSs can 
become mobile when $\eta \tau <1$ \cite{Tlidi09,TlidiEPJD2010,Tlidi12}.  
\begin{equation}
\label{speed1}
v=\frac{Q}{\tau}\sqrt{-(1-\eta \tau)},
\end{equation}
with $$Q=\sqrt{6\frac{\int_{-\infty}^{+\infty}u_{1}^{2}dxdy}{\int_{-\infty}^{+\infty}u_{2}^{2}dxdy}}$$
Due to the rotational symmetry of the generalized Swift-Hohenberg equation (\ref{SHeq}), there is no preferred direction for the 
motion of  a circularly symmetric localized structure. The instability leading to the spontaneous motion of a LS solution is a 
circle pitchfork type of bifurcation. Therefore, the $x$ axis can be chosen for the estimation of 
the velocity $\mathbf{v}$. In this case we obtain $u_{1}=\partial u_{0}(\mathbf{r})/\partial x$ and 
$u_{2}=\partial^{2}u_{0}(\mathbf{r})/\partial x^{2}$ where $u_{0}(\mathbf{r})$ is the stationary localized structure solution.
We recover the expression for the soliton velocity (\ref{speed1}) that  was obtained first in the case of the Swift-Hohenberg 
equation \cite{Tlidi09}. The speed formula (\ref{speed1}) is valid for any localized pattern, regardless of the number of peaks 
it contains. The factor $Q$ describes the spatial form of the localized pattern. This factor can be be calculated only 
 numerically. In particular, for the parameter values  $y=-0.35$, $p=-0.7$, $d=-1.2$, $a=0.75$, we obtain $Q=1.44$. The velocity 
 Eq.~(\ref{speed1}) divided by the factor $Q$ is plotted as a function the delay time for a fixed value of the feedback strength 
 as shown in Fig. \ref{velocityGSH}. The curve of the velocity has a maximum at $\tau=2/\eta$, which corresponds to the maximal 
 velocity   $v_{max}=Q\eta/2$. Note that the threshold  $\eta \tau =1$ and the expression for the formula (\ref{speed1}) have 
 been found later on for the chemical reaction-diffusion type of equations \cite{Tlidi-sonnino,Gurevich13a,GurevichPTRS}.

\begin{figure}[htbp]
\centering
\sidecaption
\includegraphics[width=0.5\textwidth]{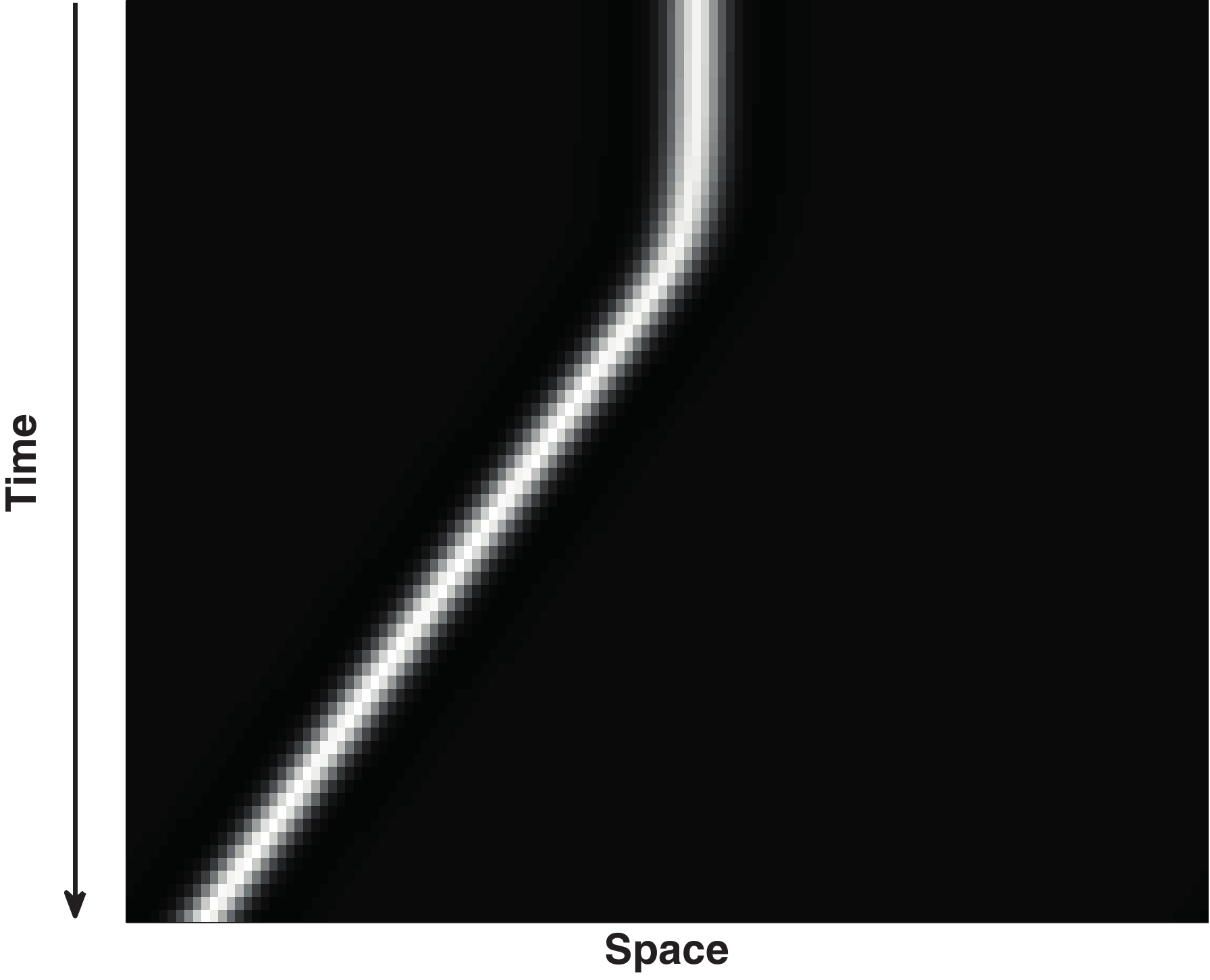}
 \caption{Space time map of a single-peaked localized structure solution to eq. (\ref{eqSHE}). Parameters are $p=-0.9$, $d=-1.5$,
$y=-0.5$, $\eta=0.15$ and $\tau=15$. Redrawn from Ref. \cite{Tinou}}
\label{mouving1D}
\end{figure}

Numerical simulations of the Eq. (\ref{SHeq}) show indeed that single and many peaked LSs exhibit a spontaneous motion as 
predicted by the above theoretical analysis, as shown in Fig. \ref{mouving1D}.

The velocity and the threshold $\eta\tau=1$ associated with the LS motion depend only on the delay parameters $\eta$ and $\tau$.
This statement is valid only in nascent optical bistability regime,where the dynamics of the system are described by the real 
order parameter equation (\ref{SHeq}), and for a fixed feedback phase of $\psi =0$ or $\psi=\pi$. In what follows we will examine
the combined role of the phase of the delay feedback $ \psi$ and the  carrier relaxation rate $\gamma$ in the framework of 
the full mean field model.
 
Since our system is isotropic, the motion of localized structures occurs in an arbitrary direction. At the pitchfork bifurcation 
the stationary LS loses stability and a branch of moving LSs with the velocity $v=|{\mathbf v}|$ bifurcates from the stationary 
LS branch of solutions. The bifurcation point can be obtained from the first order expansion of the uniformly moving LS in power 
series of the small velocity $v$. Close to the bifurcation point, the uniformly moving LS can be expanded in power series in the 
small velocity $v$ and through the solvability condition, we obtain the  drift instability threshold 
\cite{Pimenov,VladimirovPTRS}
\begin{equation}
\eta\tau =\frac{1+\gamma^{-1}(b/c) }{ \sqrt{1+(a/c)^2}\cos[\varphi+\arctan{(a/c)}]}
\label{thresholdlimit}
\end{equation}
with
\begin{eqnarray}
a&=& \langle \psi_1^{\dagger}, \psi_2 \rangle-\langle \psi_2^{\dagger}, \psi_1\rangle,\quad\\\nonumber
b&=&\langle \psi_3^{\dagger}, \psi_3 \rangle,\quad\label{coeff}\\
c&=&\langle \psi_1^{\dagger},\, \psi_1\rangle + \langle \psi_2^{\dagger},\,\psi_2\rangle, \nonumber
\end{eqnarray}
and
\begin{equation}
{\psi}=\left(\psi_1,\,\psi_2,\,\psi_3\right)^T=\partial_x\left(X_0,\,Y_0,\,N_0\right)^T \label{nm}
\end{equation}
a translational neutral mode of the linear operator $L$, $L\psi=0$, while  
${\psi}^{\dagger}=\left(\psi_1^{\dagger},\,\psi_2^{\dagger},\,\psi_3^{\dagger}\right)^T$  is the corresponding solution of the 
homogeneous adjoint problem $L^\dagger \psi^{\dagger} = 0$.  The real $X_0(x,y)$ and the imaginary  $Y_0(x,y)$ parts of the 
electric field $E_0$ and the  carrier density $N_0(x,y)$ are the stationary axially symmetric LS profiles.  They correspond to 
the time-independent solutions of Eqs.~(\ref{eq:dEdt}) and (\ref{eq:dNdt}) with $\tau=0$.  The coefficients $a$ and $b$ are 
calculated numerically using the relaxation method in two transverse dimensions. 

\begin{figure}[htbp]
\centering
\includegraphics[width=0.8\textwidth]{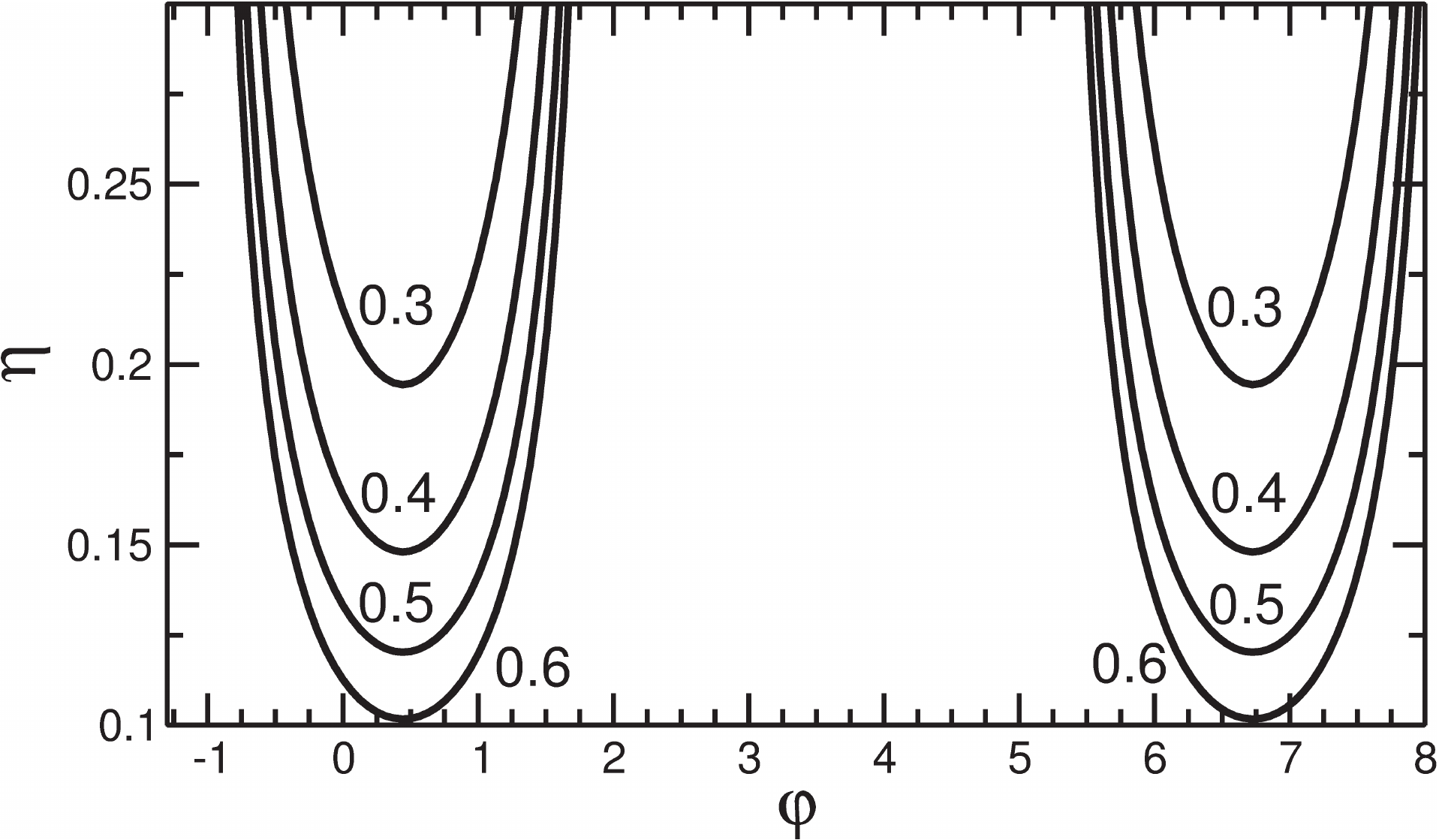}
 \caption{Threshold associated with the drift bifurcation as a function of the  feedback phase $\varphi$ is plotted for 
 different values of the carrier relaxation rate $\gamma$. Parameters are  $\mu = 1.0$, $\theta=-2.0$, $C=0.45$, 
 $\alpha=5.0$, $\tau=100$, $d = 0.052$, $E_i = 0.8$, $I = 2$. The values of the parameter $\gamma$ are shown in the figure. 
 Redrawn from \cite{VladimirovPTRS}.}
\label{ThresholdFUll}
\end{figure}

From the expression of the threshold associated with the drift instability Eq.~(\ref{thresholdlimit}), we see that the  
product $\eta \tau$ is not unity as in the case of the generalized Swift-Hohenberg equation, but depends strongly on the 
feedback phase $\varphi$ and carrier relaxation rate $\gamma$.  We plot in Fig. \ref{ThresholdFUll}, the threshold $\eta$ 
associated with the drift instability as a function of the phase of the delay feedback. The numbers on top of the different 
curves are the different values of the carrier decay rate $\gamma$. The carrier relaxation rate strongly affects the 
threshold associated with the drift instability as shown in Fig. \ref{ThresholdFUll}. When we increase the carrier relaxation 
rate, the threshold associated with the moving LS gets higher. In addition, we see from Fig. \ref{ThresholdFUll} that, whatever 
the value of  the carrier relaxation rate, the drift instability occurs only within the subinterval 
$(\varphi_{min}-\pi/2,\varphi_{min}+\pi/2)$ of the interval $(\varphi_{min}-\pi,\varphi_{min}+\pi)$, where 
$\varphi_{min}=-\arctan{a}$ is the delay feedback phase, corresponding to the lowest critical feedback rate 
$\eta_0^{min}=(1+\gamma^{-1}b)/(\tau\sqrt{1+a^2})$. Note that when $\eta_0$ increases very rapidly when approaching the 
boundaries of these subintervals.  In the limit of fast carrier response, $\gamma \gg 1$, and for zero feedback phase, 
$\varphi=0$, we recover the expression  (\ref{thresholdlimit}), the threshold formula $\eta\tau = 1$ that has been obtained in 
both variational Swift-Hohenberg equation \cite{Tlidi09}, and in a modified nonvariational one \cite{Tlidi12}.  Note that at 
$\gamma\to\infty$, $a\neq 0$, and $\varphi=-\arctan{a}$ the critical feedback rate appears to be smaller than that obtained for 
the real Swift-Hohenberg equation, $\eta\tau={(1+a^2)}^{-1/2}<1$.

\begin{figure}[htbp]
\centering
\includegraphics[width=0.8\textwidth]{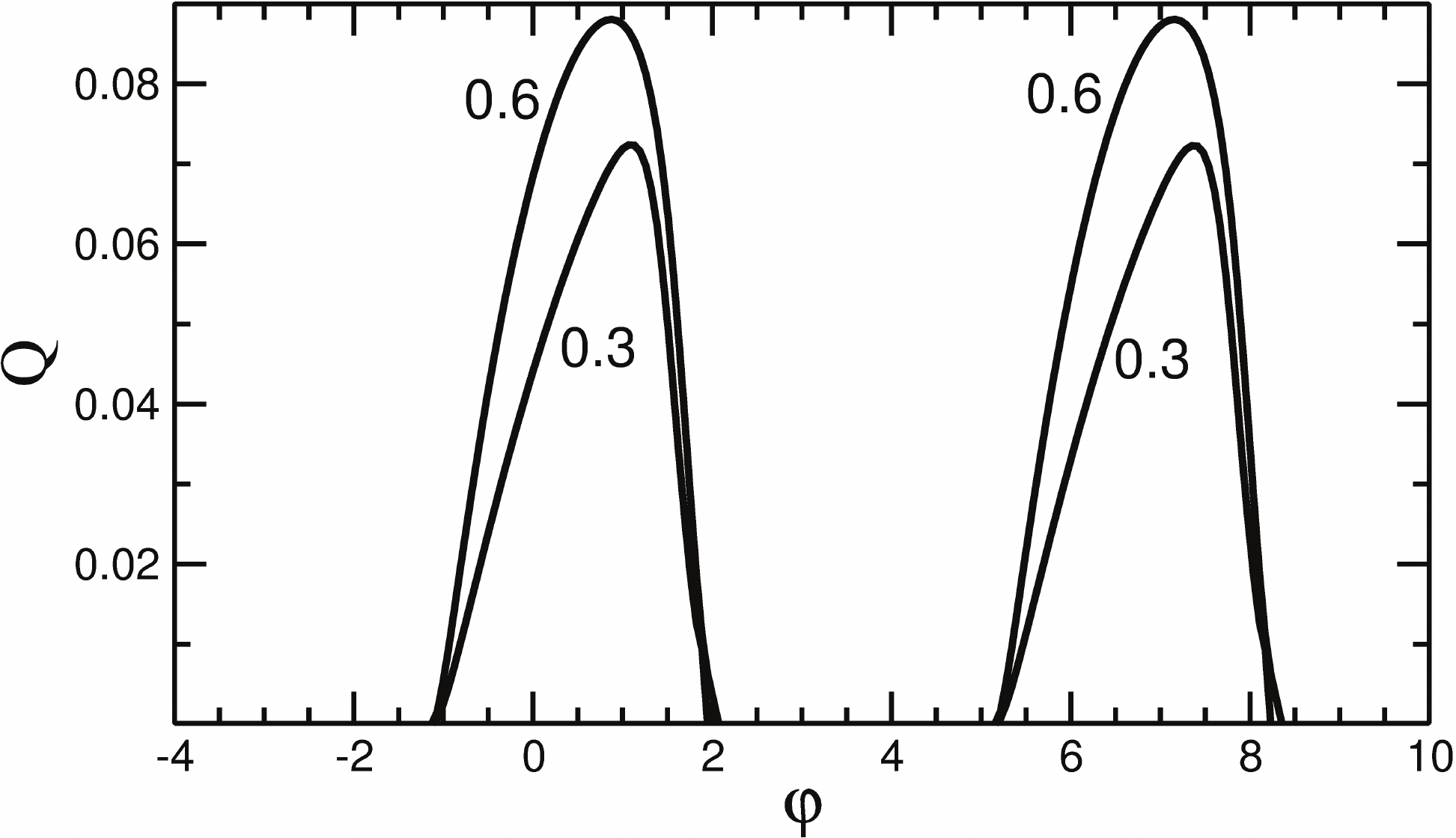}
 \caption{$Q$ factor as a function of the feedback phase $\varphi$ , for different values of the carrier relaxation rate $\gamma$.
The $Q$ factor describes the growth rate of the LS velocity with the square root of the deviation from the critical 
 feedback rate. Parameters are  $\mu = 1.0$, $\theta=-2.0$, $C=0.45$, $\alpha=5.0$, $\tau=100$, $d = 0.052$, $E_i = 0.8$, 
 $I = 2$. The values of the parameter $\gamma$ are shown in the figure.  Redrawn from \cite{VladimirovPTRS}}
\label{Qfactor}
\end{figure}

To estimate the velocity of moving localized structure, we expand the slowly moving localized solution in the limit of a small 
velocity $v$. The detailed calculations can be found in \cite{VladimirovPTRS}.  The velocity $v$ of LSs then obeys  

\begin{equation}
v=\sqrt{\delta\eta}Q \textrm{ with  } Q=(1/\tau)\sqrt{q/(r\eta)}.
\end{equation}
The factor $Q$ is important since it determines how fast the LS speed increases with the square root of the distance from the 
critical feedback rate. The $Q$ factor depends on the delay feedback phase as shown Fig. \ref{Qfactor}. The  $\delta\eta$ 
denotes the deviation of the feedback strength from the bifurcation point associated with the moving LS. The coefficients $r$ 
and $q$ are $q=a\sin{\varphi}+c\cos{\varphi}$, and $r=f\sin{\varphi}+g\cos{\varphi}+{\cal O}(\tau^{-1})$ with 
$f=\langle \psi_1^{\dagger}, \partial_{xxx} Y_0\rangle-\langle \psi_2^{\dagger}, \partial_{xxx} X_0\rangle$, 
$h=\langle \psi_3^{\dagger}, \partial_{xxx} N_0 \rangle$, $g=\langle \psi_1^{\dagger},\, \partial_{xxx} X_0\rangle + \langle 
\psi_2^{\dagger},\,\partial_{xxx} Y_0\rangle$.  The coefficients $a$ and $c$ are defined in Eq. \ref{coeff}. 

\begin{figure}[htbp]
\centering
\includegraphics[width=0.8\textwidth]{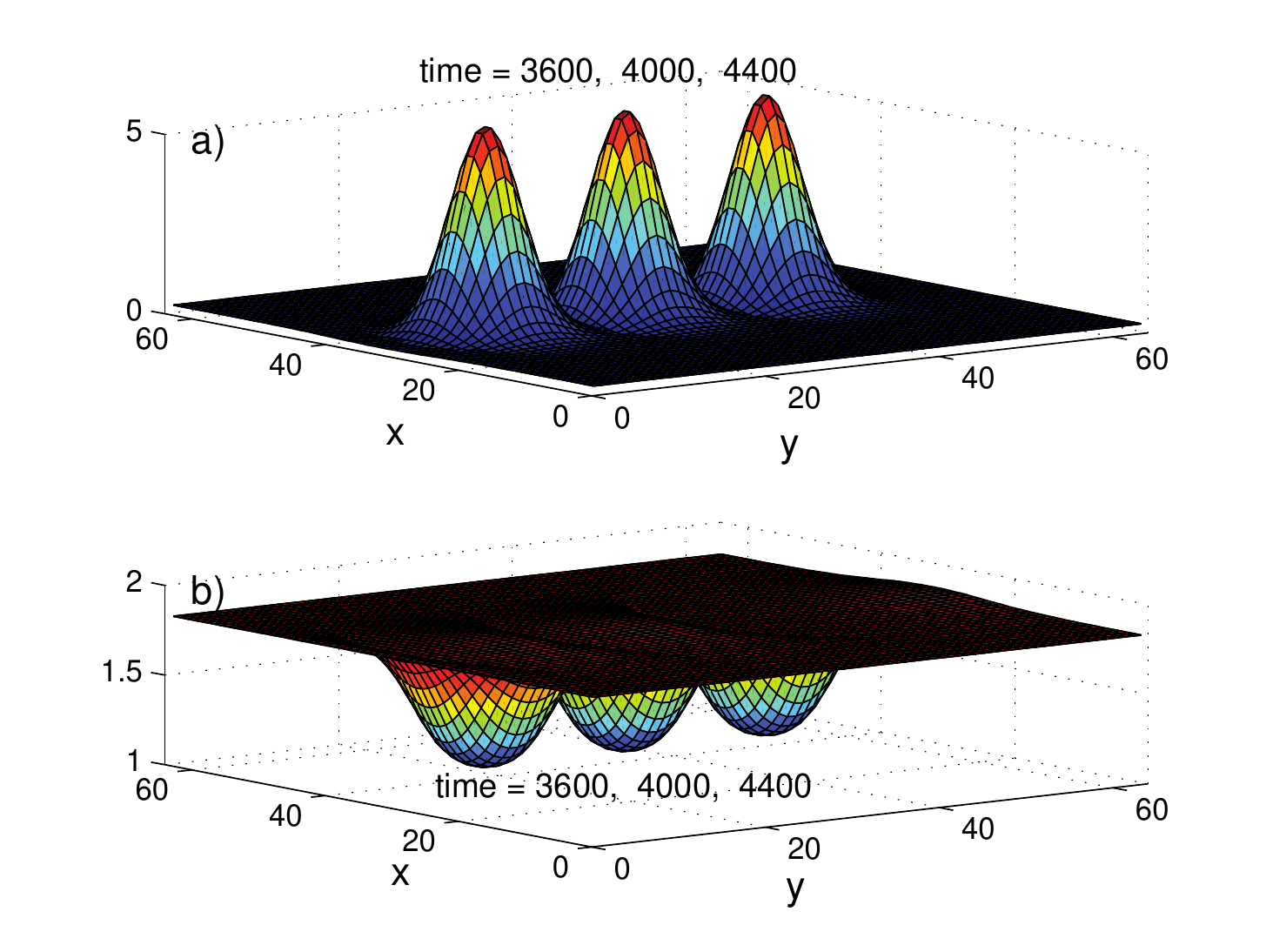}
 \caption{Field intensity $|E|^2$(top) and carrier density $N$ (bottom) of a single peaked two-dimensional moving cavity soliton
 at different times. Parameters are $C=0.45$, $\theta=-2$, $\alpha=5$, $\gamma=0.05$, $d=0.052$, $\mu=2$, $E_i=0.8$, $\tau=200$, 
 $\eta=0.07$, $\varphi=3.5$. Results obtained using Eqs. \ref{eq:dEdt} and \ref{eq:dNdt}. Redrawn from \cite{Pimenov}.}
\label{mouving2D}
\end{figure}

Numerical simulations of the full model  Eqs. (\ref{eq:dEdt},\ref{eq:dNdt}) and the generalized Swift-Hohenberg equation 
Eq. (\ref{SHeq}) agree with the above theoretical predictions. Indeed, when we choose  parameter values such as the system 
operates above the threshold associated with the motion of LS, a single-peaked LS exhibits a regular motion in the transverse 
plane of the cavity as shown in Fig. \ref{mouving1D} (1-dimensional setting, (Eq.\ref{SHeq})) and 
\ref{mouving2D}(2-dimensional setting, Eqs (\ref{eq:dEdt}) and (\ref{eq:dNdt})).

Note, however, that the motion of LS under the effects of delay feedback is not always regular. It has been shown recently that 
LSs can exhibit a temporal chaos: numerical simulations of a broad area VCSEL with saturable absorber subjected to time-delayed 
optical feedback have shown evidence of complex temporal dynamics of LSs \cite{chaoticCS}. This spatio-temporal chaos is 
localized in space. More recently,  in the absence of decay feedback, it has been shown that the VCSEL with saturable absorber 
without optical injection injection may exhibit LSs that drift and oscillate simultaneously, and  a chaotic behavior 
\cite{PratiEPJD,PratiPTRS}. 
\section{Conclusions and perspectives}
In this chapter we have investigated the formation of localized structures in a Vertical-Cavity Surface-Emitting Laser subject 
to optical injection. This device consists of a medium size bottom-emitting InGaAs multiple quantum well VCSEL operating in an 
injection locked regime. In this regime,  we have described experimentally  the formation of stationary localized structures of 
light in the transverse section of this device. The experimental part has been performed in an injection locked regime and in 
the absence of delay feedback control. We have characterized LSs by drawing their bifurcation diagram and performed a numerical 
simulation for the full model and the generalized Swift-Hohenberg equation.

Then we have described the space-time dynamics of a VCSEL by adding the delay feedback control in the modelling by adopting a 
mean-field approach. The time-delayed feedback  is modelled following a Rosanov-Lang-Kobayashi approach 
\cite{Rosanovdelay,RLang1980}. We have then analysed theoretically the effect of time delayed feedback from an external mirror 
on the stability of transverse localized structures in a broad area VCSEL. We have derived a real order parameter equation of 
the Swift-Hohenberg type with delay feedback. This analysis is only valid close the nascent optical bistability and close to 
large wavelength pattern forming regime. In this double limit, we have estimated the threshold associated with the 
drift-instability leading to the spontaneous motion of LS. Explicitly the threshold is given by a simple formula $\eta\tau=1$. 
We conclude that the condition under which transition from motionless LS to a moving one, depends only on the delay feedback 
$\eta$ and the delay time $\tau$ and not on the other dynamical parameters of the VCSEL system. This conclusion is valid for the 
variational and nonvariational Swift-Hohenberg equations, and in reaction-diffusion systems. However, in optics, the role of the 
phase is important because the intensity and the phase of the light operate on the same time scale. We have investigated in the 
last part of this chapter the  role of the phase of the feedback on the mobility of LS. We have shown that, depending on the 
phase of the feedback, it can have either stabilizing or destabilizing effect on the LSs. In particular, when the interference 
between the LS field and the feedback field is destructive, the LS can be destabilized via a pitchfork bifurcation, where a 
branch of uniformly moving LS bifurcates from the stationary one. We have calculated analytically the threshold value of the 
feedback rate corresponding to this bifurcation and demonstrated that the faster the carrier relaxation rate in the 
semiconductor medium, the lower the threshold of the spontaneous drift instability induced by the feedback. This is a generic 
and robust destabilization mechanism in one and two spatial dimensions settings and could be applied to a large class of optical 
systems under time-delay control.

We have described spatially localized structures, recent investigations have shown that temporal localized structures have been 
found in fibre resonators \cite{Leo,Tlidi-Gelens,Tlidi-Lyes,Nicolas-OE-14,Skryanbin-14,BahloulPTRS} and in VCSELS with a 
saturable absorber \cite{Marconi2014}.  On the other hand, it has been shown that the combinated influence of diffraction and 
chromatic dispersion leads to the formation of three dimensional localized structures often called light bullet  
\cite{k48-1,k48,k48-3,k48-4}. We plan in the  future to investigate the effect of delayed feedback on the spontaneous motion of 
three dimensional light bullets. 

In order to check our theoretical predictions, we also plan to investigate experimentally the formation of moving LS in VCSELs. 
In addition, we will analyse the role of local polarization dynamics in the formation of LSs in the transverse plane of the 
VCSEL. This would allow us to study the spontaneous motion of vector LSs with different polarizations under the effect of 
delayed feedback.

A.G.V. and A.P. acknowledge the support from SFB 787 of the DFG. A.G.V. acknowledges the support of the EU FP7 ITN PROPHET and 
E.T.S. Walton Visitors Award of the Science Foundation Ireland. M.T. received support from the Fonds National de la Recherche 
Scientifique (Belgium). This research was supported in part by the Interuniversity Attraction Poles program of the Belgian 
Science Policy Office under Grant No. IAP P7-35.

\end{document}